\documentclass[reqno,12pt,draft]{article}
\usepackage{amssymb,euscript}
\usepackage{amsmath,amssymb}

\newcommand{\ie}{{\it i.e.}}
\newcommand{\eg}{{\it e.g.}}
\newcommand{\e}{\begin{equation}}
\newcommand{\ee}{\end{equation}}
\newcommand{\ea}{\begin{eqnarray}}
\newcommand{\eea}{\end{eqnarray}}

\newcommand{\Tr}{{\rm Tr}}

\newcommand{\bZ}{{\bar Z}}
\newcommand{\ba}{{\bar a}}
\newcommand{\bb}{{\bar b}}
\newcommand{\bc}{{\bar c}}
\newcommand{\bd}{{\bar d}}
\newcommand{\bbe}{{\bar e}}
\newcommand{\bbf}{{\bar f}}
\newcommand{\bg}{{\bar g}}

\begin{document}

\begin{flushright}
\end{flushright}
\begin{flushright}
\end{flushright}
\begin{center}

{\LARGE {\sc Three-Algebras and ${\cal N}=6$ Chern-Simons Gauge
Theories \\ } }

\bigskip
{\sc Jonathan Bagger\footnote{bagger@jhu.edu}}\\
{Department of Physics and Astronomy\\
Johns Hopkins University\\
3400 North Charles Street\\
Baltimore, MD 21218, USA }
\\
\bigskip
and
\bigskip
\\
{\sc Neil Lambert\footnote{neil.lambert@kcl.ac.uk}} \\
{Department of Mathematics\\
King's College London\\
The Strand\\
London WC2R 2LS, UK\\}

\end{center}

\bigskip
\begin{center}
{\bf {\sc Abstract}}
\end{center}

We derive the general form for a three-dimensional scale-invariant
field theory with ${\cal N}=6$ supersymmetry, $SU(4)$
R-symmetry and a $U(1)$ global symmetry.  The results can be
written in terms of a 3-algebra in which the triple product is not
antisymmetric. For a specific choice of 3-algebra we obtain the
${\cal N}=6$ theories that have been recently proposed as models
for M2-branes in an ${\mathbb R}^8/{\mathbb Z}_k$ orbifold
background.

\newpage

\section{\sl Introduction}

M2-branes have recently been enjoying a period of considerable
interest.  One hopes that an understanding of the dynamics of
multiple M2-branes will lead to a deeper and more microscopic
understanding of M-theory.  Motivated by the papers \cite{Basu:2004ed}
and \cite{Schwarz:2004yj}, in \cite{Bagger:2006sk} we proposed a
field-theory model of multiple M2-branes. This model was shown to
admit ${\cal N}=8$ supersymmetry (16 supercharges) in \cite{Gustavsson:2007vu}
and in \cite{Bagger:2007jr}, where the Lagrangian was also given.
In this approach, the scalars and fermions take values in a 3-algebra
$\cal A$.

A 3-algebra is a vector space with basis $T^a$, $a=1,...,N$, endowed
with a triple product \cite{Bagger:2007jr},
\begin{equation}
[T^a,T^b;{\overline T}^\bc] = f^{ab\bc}{}_{d}\,T^d.
\end{equation}
Note that here we take the 3-algebra to be a {\it complex}
vector space, and we have used a slightly different
notation to keep track of the fact that, in this paper,
$[\cdot,\cdot;\cdot]$ need only be antisymmetric in
the first two indices. Furthermore, we require the
$f^{ab\bc}{}_d$ to satisfy the following
fundamental identity,
\begin{equation}\label{FI}
f^{ef\bg}{}_bf^{cb\ba}{}_d +f^{fe\ba}{}_bf^{cb\bg}{}_d+
f^{*\bg\ba f}{}_\bb f^{ce\bb}{}_d+f^{*\ba\bg e}{}_\bb f^{cf\bb}{}_d=0.
\end{equation}
(We will give an alternative characterization of this condition in
equation (\ref{newFI}) below.) In \cite{Bagger:2007jr}
(and also \cite{Gustavsson:2007vu}), we also required the
$f^{abc}{}_d$ to be real and antisymmetric in $a,b,c$. In that case,
for any such triple product, one finds equations of motion that are
invariant under 16 supersymmetries and $SO(8)$ R-symmetry.

To construct a Lagrangian we require a trace form on the 3-algebra
that is linear in the second entry and complex anti-linear in the
first. This provides an inner product,
\begin{equation}
h^{\ba b} = \Tr({\overline T}^\ba, T^b) .
\end{equation}
For $h^{ab}$ and $f^{abc}{}_d$ real, gauge invariance implies that
$f^{abcd}=f^{abc}{}_{e}h^{ed}$ is totally antisymmetric. This leads
to a Chern-Simons Lagrangian with 16 supersymmetries and $SO(8)$
R-symmetry \cite{Bagger:2007jr}.  When $h^{ab}$ is also positive
definite, it was recently shown that the one known example,
in which $f^{abcd}\propto\varepsilon^{abcd}$, is essentially
unique.\footnote{All other 3-algebras are direct sums of the minimal
four-dimensional 3-algebra
\cite{nagy}--\cite{Papadopoulos:2008sk}.} In
\cite{Lambert:2008et,Distler:2008mk} this maximally supersymmetric
field theory was identified as describing two M2-branes in an
${\mathbb R}^8/{\mathbb Z}_2$ orbifold background.

Recently, there have been several attempts to relax these
assumptions and construct additional models. In \cite{Gran:2008vi}
it was suggested that $f^{abc}{}_d$ need not be totally
antisymmetric, just antisymmetric $a,b,c$, and indeed this leads to
an infinite number of models using the 3-algebra given in
\cite{Awata:1999dz}. The equations of motion of \cite{Bagger:2007jr}
are still invariant under the 16 supersymmetries, but there is no
gauge-invariant metric so it is not clear how to construct physical
quantities such as energy.

More recently there have been proposals in which the metric $h^{ab}$
has a Lorentzian signature \cite{Gomis:2008uv}--\cite{Ho:2008ei}.
This allows one to construct an associated 3-algebra for any Lie
algebra, and the corresponding ${\cal N}=8$ Lagrangian
\cite{Bagger:2007jr} has been proposed  to describe M2-branes in
flat ${\mathbb R}^8$ \cite{Gomis:2008uv}--\cite{Ho:2008ei}. Although
these models are built on a 3-algebra without a positive definite
norm, the corresponding quantum theories have been argued to be
unitary \cite{Gomis:2008uv}--\cite{Ho:2008ei} and there are some
encouraging features \cite{Lin:2008qp}--\cite{Cecotti:2008qs}. The
current status of these models is unclear.  In particular, one
method for removing the negative norm states leads back to maximally
supersymmetric Yang-Mills theory
\cite{Bandres:2008kj,Ezhuthachan:2008ch}, although in a form that
possesses both $SO(8)$ and spontaneously broken  conformal symmetry.

Another option is to look for theories with a reduced number of
supersymmetries.  In \cite{Gaiotto:2008sd}--\cite{Hosomichi:2008jd}
a class of Chern-Simons Lagrangians with ${\cal N}=4$ supersymmetry
(8 supercharges) was constructed.  More recently, in
\cite{Aharony:2008ug} an infinite class of brane configurations was
given whose low energy effective Lagrangian is a Chern-Simons theory
with $SO(6)$ R-symmetry and ${\cal N}=6$ supersymmetry (12
supercharges).  These theories are related to $N$ M2-branes in
${\mathbb R}^8/{\mathbb Z}_k$, including $k=1$. The Lagrangians were
studied in detail in \cite{Benna:2008zy}--\cite{Grignani:2008is}.
More theories with ${\cal N}=5$ and ${\cal N}=6$ have also recently
appeared in \cite{Hosomichi:2008jb}.

Thus it is of interest to generalize the construction of our model,
based on 3-algebras, to the case of ${\cal N}=6$ supersymmetry. We
will see that this can be accomplished by relaxing the conditions on
the triple product so that it is no longer real and antisymmetric
in all three indices. Rather it is required to satisfy
\begin{equation}\label{Riemann}
f^{ab\bc\bd}=-f^{ba\bc\bd} \qquad{\rm and}\qquad f^{ab\bc\bd}= {f}^{*\bc\bd ab} .
\end{equation}
The triple product is also required to satisfy the fundamental identity
(\ref{FI}).

The rest of this paper is organized as follows.  In section 2 we
revisit the analysis of  \cite{Bagger:2007jr}, trying to be as
general as possible. We will see that the model presented there is
the most general with ${\cal N}=8$ supersymmetry, scale invariance
and $SO(8)$ R-symmetry. Section 3 contains the main results of this
paper.  We follow the construction of \cite{Bagger:2007jr}, but only
impose ${\cal N}=6$ supersymmetry, scale invariance, $SU(4)$
R-symmetry, and a global $U(1)$.  We find the supersymmetry
transformations, the invariant Lagrangian, and the conditions on
the structure constants $f^{ab\bc\bd}$. In section 4 we discuss the
associated 3-algebra and show that a specific choice of triple
product leads directly to the models in \cite{Aharony:2008ug}, as
presented in \cite{Benna:2008zy}.  As a result, we are able to
provide the complete expressions for the Lagrangians in
\cite{Aharony:2008ug}, including all the supersymmetry
transformations, in a manifestly $SU(4)$ covariant form (see also
\cite{Gaiotto:2008cg, Hosomichi:2008jb}). Section 5 contains our conclusions.
We collect our spinor conventions and some useful identities in an
appendix.

\section{\sl ${\cal N}=8$}

Before presenting the main results of this paper, we re-examine
the closure of the $N=8$ supersymmetry transformations given in
\cite{Bagger:2007jr} (see also \cite{Gustavsson:2007vu}).  In
particular, we relax as many assumptions as possible to
find the minimum requirements on $f^{abcd}$.  We proceed
by assuming scale invariance and an $SO(8)$ R-symmetry. The
most general form for the supersymmetry
transformations is then
\begin{eqnarray}\label{susygauged}
\nonumber \delta X^I_d &=& i\bar\epsilon\Gamma^I\Psi_d\\
\nonumber \delta \Psi_d &=& D_\mu X^I_d\Gamma^\mu \Gamma^I\epsilon
-\frac{1}{6} X^I_aX^J_bX^K_c f^{abc}{}_{d}\Gamma^{IJK}\epsilon
+\frac{1}{2}X^J_aX^J_bX^I_c g^{abc}{}_d \Gamma^I\epsilon \\
\delta\tilde A_{\mu}{}^c{}_d &=& i\bar\epsilon
\Gamma_\mu\Gamma_IX^I_a\Psi_b h^{abc}{}_{d},
\end{eqnarray}
where $D_\mu$ is a covariant derivative, and $g^{abc}{}_d$ and
$h^{abc}{}_d$ define triple products on the algebra that are not
antisymmetric (a possibility that was mentioned in
\cite{Bagger:2006sk}). Without loss of generality we may assume that
$f^{abc}{}_d$ is antisymmetric in $a,b,c$, while $g^{abc}{}_d$ is
symmetric in $a,b$.  All quantities are taken to be real.

To begin we consider the closure on $X^I_d$,
\begin{eqnarray}
[\delta_1,\delta_2] X^I_d &=&  v^\mu D_\mu X^I_d
+\tilde\Lambda^c{}_d X^I_c + \Omega^{IJc}{}_dX^J_c,
\end{eqnarray}
where
\begin{eqnarray}
\nonumber v^\mu &=& -2i\bar\epsilon_2
\Gamma^\mu\epsilon_1 \\
\nonumber   \tilde\Lambda^c{}_d  &=& -i\bar\epsilon_2
\Gamma_{JK}\epsilon_1 X^J_aX^K_bf^{abc}{}_d\\
\Omega^{IJc}{}_d &=& i\bar\epsilon_2
\Gamma^{IJ}\epsilon_1 X^K_aX^K_bg^{abc}{}_d.
\end{eqnarray}
The first two terms are familiar from \cite{Bagger:2007jr}. The last
transformation, however, mixes an internal symmetry with an
R-symmetry, although we note it becomes a pure R-symmetry if
$g^{abc}{}_d$ takes the form
\begin{equation}\label{grestriction}
g^{abc}{}_d = k^{ab}\delta^c{}_d .
\end{equation}
This implies that R-symmetry must be gauged.  A similar extension
was successfully used in \cite{Gomis:2008cv,Hosomichi:2008qk}
except that the additional term was linear in $X^I$.  As a result, the
R-symmetry was not gauged, and the theory described a mass
deformation that preserved all supersymmetries but broke the
R-symmetry to $SO(4)\times SO(4)$.

R-symmetries cannot be gauged in rigid supersymmetry because
the supercharges rotate into each other (by definition)
and hence would have to become local symmetries.\footnote{We
thank J.~Maldacena for this point.}  Thus we are forced to set
$g^{abc}{}_d=0$.

We now consider the fermions.  Evaluating
$[\delta_1,\delta_2]\Psi_d$ and using the Feirz identity
(see the appendix), we find four terms involving
$\bar\epsilon_2\Gamma_{\mu}\Gamma_{LMNP}\epsilon_1$.
After some manipulations, we reduce these terms to
\begin{equation}
\label{close2}
\bar\epsilon_1\Gamma_{\nu}\Gamma_{LMNP}\epsilon_2
\Gamma^\nu\Gamma^I\Gamma^{LMNP}
\Gamma^JX^I_cX^J_a(f^{abc}{}_d-h^{abc}{}_{d})\Psi_b.
\end{equation}
Closure implies that (\ref{close2}) must vanish and hence
\begin{equation}\label{fghk}
h^{abc}{}_d = f^{abc}{}_d.
\end{equation}
Thus we are left with just one tensor $f^{abc}{}_d$.  As in
\cite{Bagger:2007jr}, the algebra closes on the fermions using
the on-shell condition
\begin{equation}
 \Gamma^\mu D_\mu\Psi_d +\frac{1}{2}\Gamma_{IJ}X^I_aX^J_b\Psi_c
f^{abc}{}_d =0.
\end{equation}

Next we turn to $[\delta_1,\delta_2]\tilde A_\mu{}^c{}_d$. Here we
find a term that is fourth order in the scalars:
\begin{equation}
(\bar\epsilon_2\Gamma_{\mu}\Gamma_{IJKL}\epsilon_1)
X^I_a X^J_eX^K_fX^L_gf^{efg}{}_{b}f^{abc}{}_{d}.
\end{equation}
This term vanishes provided that
\begin{equation}\label{otherFI}
f^{[abc}{}_{e}f^{d]ef}{}_{g}=0 .
\end{equation}
Given the antisymmetry of $f^{abc}{}_d$ in $a,b,c$, this is
equivalent to the fundamental identity (\ref{FI}). Continuing, we
find
\begin{eqnarray}
  \nonumber [\delta_1,\delta_2]\tilde A_\mu{}^c{}_d &=&  2i(\bar\epsilon_2\Gamma^\nu\epsilon_1)
  \epsilon_{\mu\nu\lambda} (X^I_aD^\lambda X^I_b + \frac{i}{2}\bar\Psi_a\Gamma^\lambda\Psi_b)f^{abc}{}_{d} \\
   && -\ 2i(\epsilon_2\Gamma_{IJ}\epsilon_1)X^I_aD_\mu X^J_bf^{abc}{}_{d}.
\end{eqnarray}
Gauge invariance requires that the last line be equal to $D_\mu \tilde
\Lambda^c{}_d$. Writing $\tilde A_\mu{}^c{}_d = f^{abc}{}_dA_{\mu ab}$,
this implies the condition
\begin{equation}
f^{abc}{}_ef^{fge}{}_d=f^{fga}{}{}_ef^{ebc}{}_d+f^{fgb}{}{}_ef^{aec}{}_d+f^{fgc}{}{}_ef^{abe}{}_d ,
\label{deriv8}
\end{equation}
which ensures that the gauge symmetry acts as a derivation.
Equation (\ref{deriv8}) is
equivalent to (\ref{otherFI}) (\eg\ see \cite{Gran:2008vi}), so we
have recovered all the ingredients of \cite{Bagger:2007jr}.

\section{\sl ${\cal N}=6$ }

In this section we relax the constraints on $f^{ab\bc}{}_d$ to construct
an infinite class of theories with fewer supersymmetries. We will
construct a Lagrangian with 12 supercharges (${\cal N}=6$
supersymmetry), $SU(4)$ R-symmetry, and a $U(1)$ internal symmetry.
We continue to assume that $h^{\ba b}$ is positive definite,
although no substantial changes arise if $h^{\ba b}$ has a different
signature.

We use a complex notation in which the $SO(8)$ R-symmetry of the
${\cal N}=8$ theory is broken to $SU(4)\times U(1)$.   The
supercharges transform under the $SU(4)$ R-symmetry; the $U(1)$
provides an additional global symmetry.  We introduce four
complex 3-algebra valued scalar fields $Z^A_a$, $A=1,2,3,4$, as well
as their complex conjugates $\bZ_{A\ba}$.  Similarly, we denote the
fermions by $\psi_{Aa}$ and their complex conjugates by $\psi^A_\ba$.
A raised $A$ index indicates that the field
is in the $\bf 4$ of $SU(4)$; a lowered index transforms in the
$\bar{\bf 4}$. We assign $Z^A_a$ and $\psi_{Aa}$ a $U(1)$ charge of
1. Complex conjugation raises or lowers the $A$ index, flips
the sign of the $U(1)$ charge, and interchanges $a \leftrightarrow \bar a$.
The supersymmetry generators $\epsilon_{AB}$ are in the $\bf 6$ of
$SU(4)$ with vanishing $U(1)$ charge.  They satisfy the reality
condition $\epsilon^{AB} = \frac{1}{2}
\varepsilon^{ABCD}\epsilon_{CD}$.

We postulate the following supersymmetry transformations
(our spinor conventions are listed in the appendix):
\begin{eqnarray}\label{susy}
\nonumber  \delta Z^A_d &=& i\bar\epsilon^{AB}\psi_{Bd} \\
 \nonumber   \delta \psi_{Bd} &=& \gamma^\mu D_\mu Z^A_d\epsilon_{AB} +
  f_1^{a\bb c}{}_dZ^C_a \bZ_{C\bb}Z^A_c \epsilon_{AB}+f_2^{ab\bc}{}_d
  Z^C_a Z^D_{b} \bZ_{B\bc}\epsilon_{CD} \\
 \delta \tilde A_\mu{}^c{}_d &=&
i\bar\epsilon_{AB}\gamma_\mu Z^A_a\psi^B_\bb f_3^{a\bb c}{}_d +
i\bar\epsilon^{AB}\gamma_\mu \bZ_{A\ba}\psi_{Bb} f_4^{\ba bc}{}_d,
\end{eqnarray}
where $f_1^{a\bb c}{}_d$, $f_2^{ab\bc}{}_d$, $f_3^{a\bb c}{}_d$ and
$f_4^{\ba bc}{}_d$ are tensors on the 3-algebra.  Without loss of
generality, we assume that $f_2^{(ab)\bc}{}_d=0$.  The covariant
derivative is defined by $D_\mu Z^A_d =
\partial_\mu Z^A_d -\tilde A_\mu{}^c{}_d Z^A_c$. Therefore we require that $D_\mu
\bZ_{A\bd}=\partial_\mu \bZ_{A\bd} - \tilde A^*_\mu{}^{\bc}{}_\bd \bZ_{A\bc}$.
Supersymmetry then requires that $D_\mu \psi^A_\bd =
\partial_\mu \psi^A_\bd -\tilde A^*_\mu{}^{\bc}{}_\bd \psi^A_\bc$ and
$D_\mu \psi_{Ad} = \partial_\mu \psi_{Ad} -\tilde
A_\mu{}^c{}_d \psi_{Ac}$. These are the most general transformations
that preserve the $SU(4)$, $U(1)$ and conformal symmetries.

In \cite{Bagger:2006sk} the ${\cal N}=8$ theory (without gauge
fields) was written in terms of such a complex notation with manifest
$SU(4)\times U(1)$ symmetry.  However, the supersymmetries
$\epsilon_{AB}$ were not considered in detail; the discussion focused
on the other four supersymmetry generators
$\varepsilon$ that are $SU(4)$ singlets with $U(1)$ charge $\pm2$.
These supersymmetries have a natural $N=2$ superspace interpretation;
they require that $f^{abcd}$ be real and totally antisymmetric.  These
supersymmetries will not, in general, be preserved in the models
presented here.  Indeed, imposing these as supersymmetries leads
to the original ${\cal N}=8$ theory (written in complex notation).

To begin, we first consider the closure of (\ref{susy}) on the
scalars. Using the identities listed in the appendix, we find that
$[\delta_1,\delta_2]Z^A_d$ only closes onto translations and a
gauge symmetry if
\begin{equation}
f_1^{a\bb c}{}_d = f_2^{ac\bb}{}_d.
\label{f1}
\end{equation}
In this case we find
\begin{equation}
\label{close3}
[\delta_1,\delta_2]Z^A_d = v^\mu D_\mu Z^A_d +
\Lambda_{\bc b}f_2^{ab\bc}{}_dZ^A_a,
\end{equation}
where
\begin{equation}\label{symgen}
v^\mu = \frac{i}{2}\bar\epsilon_2^{CD}\gamma^\mu\epsilon_{1CD},\qquad
\Lambda_{\bc b}=i(\bar\epsilon^{DE}_2\epsilon_{1CE}-
\bar\epsilon^{DE}_1\epsilon_{2CE})\bZ_{D\bc}Z^C_b.\\
\end{equation}
The second term in (\ref{close3}) is a gauge transformation:
$\delta_\Lambda Z^A_d = \Lambda_{\bc b}f_2^{ab\bc}{}_dZ^A_a$.

Next we examine the closure of the algebra on the fermions. After
some work, we find that if
\begin{equation}
f_4^{\ba bc}{}_d = -f_3^{b\ba c}{}_d \ ,
\label{f2}
\end{equation}
and
\begin{equation}
f_3^{a\bb c}{}_d =f_2^{ac\bb}{}_d\ ,
\label{f3}
\end{equation}
then
\begin{eqnarray}
\nonumber [\delta_1,\delta_2]\psi_{Dd} &=& v^\mu D_\mu \psi_{Dd} +
\Lambda_{\bb a}f_2^{ca\bb}{}_d\psi_{Dc}\\
\nonumber &&-\frac{i}{2}(\bar\epsilon_1^{AC}\epsilon_{2AD}-\bar\epsilon_2^{AC}\epsilon_{1AD})E_{Cd}\\
 &&
+\frac{i}{4}(\bar\epsilon^{AB}_1\gamma_\nu\epsilon_{2AB})\gamma^\nu
E_{Dd},
\end{eqnarray}
where
\begin{equation}
E_{Cd} = \gamma^\mu D_\mu\psi_{Cd} +f_2^{ab\bc}{}_d \psi_{Ca}
Z^D_b\bZ_{D\bc}-2f_2^{ab\bc}{}_d\psi_{Da}Z^D_b\bZ_{C\bc}-\varepsilon_{CDEF}f_2^{ab\bc}{}_d\psi^D_\bc Z^E_aZ^F_b.
\end{equation}
Thus we see that the supersymmetry algebra closes if we impose the
on-shell condition $E_{Cd}=0$.

Finally we look at the gauge field $\tilde A_\mu{}^c{}_d$.
In the closure there is
a term that is fourth order in the scalars that vanishes when
$f_2^{ab\bc}{}_d$ satisfies the fundamental identity (\ref{FI}). At
quadratic order in the fields, closure of the supersymmetry
transformations gives
\begin{eqnarray}\label{Aclose}
[\delta_1,\delta_2]\tilde A_\mu{}^c{}_d &=& -f_3^{a\bb c}{}_d
D_\mu(\Lambda_{\bb a})\\
\nonumber &&+\varepsilon_{\mu\nu\lambda}v^\nu \left(D^\lambda Z^A_a
\bZ_{A\bb}- Z^A_aD^\lambda \bZ_{A\bb}
-i\bar\psi^A_\bb\gamma^\lambda\psi_{Aa}\right)f_3^{a\bb c}{}_d .
\end{eqnarray}
Thus if we impose the on-shell condition
\begin{equation}
\tilde F_{\mu\nu}{}^c{}_d =-
\varepsilon_{\mu\nu\lambda}\left(D^\lambda Z^A_a \bZ_{A\bb}-
Z^A_aD^\lambda \bZ_{A\bb}
-i\bar\psi^A_{\bb}\gamma^\lambda\psi_{Aa}\right)f_2^{ca\bb}{}_d,
\end{equation}
we see that the supersymmetry algebra closes onto translations and
gauge transformations
\begin{equation}
[\delta_1,\delta_2]\tilde A_\mu{}^c{}_d = v^\nu \tilde
F_{\mu\nu}{}^c{}_d+D_\mu(\Lambda_{\bb a}f_2^{ca\bb}{}_d),
\end{equation}
provided that $D_\mu (f_2^{ca\bb}{}_d)=0$. This is just the statement
that $f_2^{ca\bb}{}_d$ is an invariant tensor of the gauge algebra. In
general it provides an additional condition on $f_2^{ca\bb}{}_d$.
However we will see that it follows directly from the fundamental
identity whenever there is a Lagrangian.

Let us summarize our results so far.  Henceforth we drop the
subscript $2$ on $f_2^{ab\bc}{}_d$, which we take to be an invariant
tensor of the gauge algebra that satisfies (\ref{FI}); the remaining
tensors $f_1^{a\bc b}{}_d$, $f_3^{a\bc b}{}_d$ and $f_4^{\bc ab}{}_d$ are
related to $f^{ab\bc}{}_d$ through (\ref{f1}), (\ref{f2}) and
(\ref{f3}). The supersymmetry transformations are
\begin{eqnarray}\label{finalsusy}
\nonumber  \delta Z^A_d &=& i\bar\epsilon^{AB}\psi_{Bd} \\
\nonumber  \delta \psi_{Bd} &=& \gamma^\mu D_\mu Z^A_d\epsilon_{AB} +
  f^{ab\bc}{}_dZ^C_a Z^A_b \bZ_{C\bc} \epsilon_{AB}+
  f^{ab\bc}{}_d Z^C_a Z^D_{b} \bZ_{B\bc}\epsilon_{CD} \\
  \delta \tilde A_\mu{}^c{}_d &=&
-i\bar\epsilon_{AB}\gamma_\mu Z^A_a\psi^B_\bb f^{ca\bb}{}_d +
i\bar\epsilon^{AB}\gamma_\mu \bZ_{A\bb}\psi_{Ba} f^{ca\bb}{}_d .
\end{eqnarray}
In the case that $f^{abcd}$
is real and antisymmetric in $a,b,c$, we recover the supersymmetry
transformations of the ${\cal N}=8$ theory.

Let us now construct an invariant Lagrangian.  We have seen that
the supersymmetry algebra closes into a translation plus a gauge
transformation.  On the field $\bZ_{A\bd}$, we find
\begin{equation}
[\delta_1,\delta_2]\bZ_{A\bd} = v^\mu D_\mu \bZ_{A\bd} +
\Lambda^*_{c \bb }f^{*\ba\bb c}{}_\bd\bZ_{A\ba},
\label{cl}
\end{equation}
with $v$ and $\Lambda_{\bc b}$ given in (\ref{symgen}).  The second
term is a gauge transformation, $\delta_\Lambda \bZ_{A\bd} =
\Lambda^*_{c \bb} f^{*\ba\bb c}{}_\bd \bZ_{A\ba} =
-\Lambda_{\bb c} f^{*\ba\bb c}{}_\bd \bZ_{A\ba}$.  To construct a
gauge-invariant Lagrangian (or, for that matter, any gauge-invariant
observable) we need the metric to be gauge invariant, namely
$\delta_\Lambda (h^{\ba b} \bZ_{A\ba} Z^A_b)=0$.  Therefore we
must require
\begin{equation}\label{last}
f^{ab\bc\bd} = {f^*}^{\bc\bd ab},
\end{equation}
where $f^{ab\bc\bd} =f^{ab\bc}{}_e h^{\bd e}$.  This implies that
$(\tilde\Lambda^{c \bd})^* = -\tilde \Lambda^{d \bc}$, where
\begin{equation}
\tilde \Lambda^{c \bd} = \Lambda_{\bb a}f^{ca\bb \bd},
\end{equation}
so the transformation parameters $\tilde \Lambda^c{}_d$
are elements of $u(N)$, although they are not in general all of $u(N)$.

The first term in (\ref{cl}) contains the translation.  Note that it
appears as part of a covariant derivative, $v^\mu D_\mu
\bZ_{A\bd} = v^\mu\partial_\mu \bZ_{A\bd} - v^\mu \tilde A^*_\mu{}^\bc{}_\bd\bZ_{A\bc}$
The first part is the translation, while the second is another gauge
transformation, with parameter $\tilde \Lambda^{*\bc}{}_\bd=-v^\mu
\tilde A^*_\mu{}^\bc{}_\bd$.  This implies that the gauge field also takes
values in $u(N)$.

With these results, it is not hard to show that an invariant
Lagrangian (up to boundary terms) is given by
\begin{eqnarray}\label{lagrangian}
\nonumber {\cal L} &=& - D^\mu \bZ_A^a D_\mu Z^A_a -
i\bar\psi^{Aa}\gamma^\mu D_\mu\psi_{Aa} -V+{\cal L}_{CS}\\[2mm]
&& -i f^{ab\bc\bd}\bar\psi^A_\bd \psi_{Aa}
Z^B_b\bZ_{B\bc}+2if^{ab\bc\bd}\bar\psi^A_\bd\psi_{Ba}Z^B_b\bZ_{A\bc}\\
\nonumber
&&+\frac{i}{2}\varepsilon_{ABCD}f^{ab\bc\bd}\bar\psi^A_\bd\psi^B_\bc Z^C_aZ^D_b
-\frac{i}{2}\varepsilon^{ABCD}f^{cd\ba\bb}\bar\psi_{Ac}\psi_{Bd}\bZ_{C\ba}\bZ_{D\bb}\
,
\end{eqnarray}
where the potential is
\begin{equation}
V = \frac{2}{3}\Upsilon^{CD}_{Bd}\bar\Upsilon_{CD}^{Bd} ,
\end{equation}
where
\begin{equation}
\Upsilon^{CD}_{Bd} = f^{ab\bc}{}_dZ^C_aZ^D_b\bZ_{B\bc}
-\frac{1}{2}\delta^C_Bf^{ab\bc}{}_dZ^E_aZ^D_b\bZ_{E\bc}+\frac{1}{2}\delta^D_Bf^{ab\bc}{}_dZ^E_aZ^C_b\bZ_{E\bc}.
\end{equation}
The zero-energy solutions correspond to $\Upsilon^{CD}_{Bd}=0$.  This
is equivalent to $\Upsilon^{CD}_{Bd}\epsilon_{CD}= 0$ for arbitrary
$\epsilon_{CD}$, which implies that the zero-energy solutions preserve
all 12 supersymmetries.

The `twisted' Chern-Simons term ${\cal L}_{CS}$ is given by
\begin{equation}
{\cal
L}_{CS}=\frac{1}{2}\varepsilon^{\mu\nu\lambda}\left(f^{ab\bc\bd}A_{\mu
\bc b}\partial_\nu A_{\lambda \bd a} +\frac{2}{3}f^{ac\bd}{}_gf^{ge\bbf\bb}
A_{\mu \bb a}A_{\nu \bd c}A_{\lambda \bbf e}\right).
\end{equation}
It satisfies
\begin{equation}
\frac{\delta{\cal L}_{CS}}{\delta \tilde A_{\lambda}{}^{a\bb}}f^{ac\bd\bb} =
\frac{1}{2}\varepsilon^{\lambda\mu\nu}\tilde F_{\mu\nu}{}^{c\bd},
\end{equation}
up to integration by parts,
where $\tilde F_{\mu\nu}{}^a{}_b = -\partial_\mu \tilde
A_\nu{}^a{}_b+\partial_\nu \tilde A_\mu{}^a{}_b +  \tilde
A_\nu{}^a{}_e\tilde A_\mu{}^e{}_b- \tilde A_\mu{}^a{}_e\tilde
A_\nu{}^e{}_b $.  Just as in \cite{Bagger:2007jr}, this term can be
viewed as a function of $\tilde A_\mu{}^c{}_d$ and not $A_{\mu c\bd}$.

Note that the Lagrangian (\ref{lagrangian}) is automatically gauge
invariant since it is supersymmetric and supersymmetries close into
gauge transformations. One can also confirm that the equations of
motion give the on-shell conditions that we
found above for closure of the supersymmetry algebra.

\section{\sl Three-Algebras and Their Construction}

A given tensor $f^{ab\bc}{}_d$ defines a triple product on the
algebra with (complex) generators $T^a$:
\begin{equation}
[T^a,T^b;\overline{T}^c] = f^{ab\bc}{}_dT^d,
\end{equation}
which is linear and anti-symmetric in the first two entries and
complex anti-linear in the third. In a sense one may think of
$[\cdot\ ,\cdot\ ;\cdot\ ]$ as generating a map from the 3-algebra
$\cal A$ into the space of endomorphisms of $\cal A$, \ie\ for a
fixed pair $Y,\overline{Z}\in {\cal A}$, $[\cdot,Y;\overline{Z}]$
defines a linear map of $\cal A$ into itself. We then obtain a
triple product of any three elements $X,Y,\overline{Z}\in \cal A$ by
evaluating the map $[\cdot,Y;\overline{Z}]$ on $X$.

For the case at hand, the triple product generates a gauge symmetry
\begin{equation}
\delta Z^A_d = \Lambda_{\bb a}f^{ca\bb}{}_d Z^A_c\ .
\end{equation}
This is similar to the gauge symmetry in \cite{Bagger:2007jr}, but
there are some important differences. Let us generalize the
discussion of \cite{Bagger:2007vi}.  In what follows, we assume
the existence of a gauge-invariant metric, so $\Lambda^a{}_b$
extracted from (\ref{symgen}) is an element of $u(N)$. The symmetries
(\ref{Riemann}) imply that $\tilde \Lambda^{c}{}_{d} = f^{ca\bb}{}_{d}
\Lambda_{\bb a}$ is also an element of $u(N)$ (where we assume for
concreteness that the metric is positive definite). Thus $f^{ca\bb}{}_{d}$
defines a map $f$: $u(N) \to u(N)$;
\begin{equation}
f(\Lambda)^{c}{}_{d}= \Lambda_{\bb a}f^{ca\bb}{}_{d}\ .
\end{equation}

Let $\cal G$ be the vector space generated by the image of $f$. The
fundamental identity (\ref{FI}) implies that
\begin{equation}
[f(\Lambda_1),f(\Lambda_2)] = f(\Lambda_3)
\end{equation}
where $\Lambda_{3\ba b} = \Lambda_{1 \ba e}\Lambda_{2\bg f} f^{ef\bg}{}_b
-\Lambda_{1 \bbe b}\Lambda_{2\bg f} 
f^{*\bbe\bg f}{}_\ba$.
In other words, the space $\cal G$ of gauge
transformations is closed under the ordinary matrix commutator and
is therefore a Lie subalgebra of $u(N)$.  In the special case that
$f^{abcd} = -f^{acbd}$, we see that $f^{abcd}$ is real and totally
antisymmetric. In that case $\cal G$ is generated by antisymmetric
elements of $u(N)$. These are necessarily real and hence we recover
the construction of \cite{Bagger:2007jr} in which $\cal G$ is a Lie
subalgebra of $so(N)$.

Using the metric and the  condition (\ref{last}), we write
the fundamental identity (\ref{FI}) as
\begin{equation}\label{newFI} f^{ab\bc}{}_ef^{ef\bg}{}_d = f^{af\bg}{}_ef^{eb\bc}{}_d +
f^{bf\bg}{}_ef^{ae\bc}{}_d - f_\bbe{}^{f\bg\bc}f^{ab\bbe}{}_d ,
\end{equation}
which says that the gauge symmetry acts as a
derivation. In particular if we contract (\ref{newFI}) with
$\Lambda_{\bg f}$ it is equivalent to the condition
\begin{equation}\label{derivation}
\delta [Z^A,Z^B;\bar Z_C] = [\delta Z^A,Z^B;\bar Z_C] + [Z^A,\delta
Z^B;\bar Z_C] + [Z^A,Z^B;\delta \bar Z_C].
\end{equation}
where  $\delta Z^A = \tilde\Lambda^a{}_b Z^A_aT^b$. Thus we see that
the gauge symmetry acts as a derivation.

To continue we give a characterization of tensors $f^{ab\bc\bd}$ that
satisfy (\ref{FI}) and (\ref{Riemann}) by adapting a discussion from
\cite{Gustavsson:2008dy}.  As we have noted, $f^{ab\bc\bd}$ generates
the Lie algebra  $\cal G$ of gauge transformations.  For any two
generators $T^a$ and $T^b$, we write
\begin{equation}
[X,T^a;\overline{T}^\bb ]_d= \Gamma_A^{a\bb}(t^A)^c{}_d X_c,
\end{equation}
where the $\Gamma^{a\bb}_A$ are constants and the $t^A$ are a matrix
representation of $\cal G$ inside $u(N)$. In particular, the $t^A$
are anti-Hermitian. We note that
\begin{equation}\label{fis}
f^{ab\bc\bd} = {\rm Tr}(\overline{T}^\bd,[T^a,T^b;\overline{T}^\bc]) ,
\end{equation}
and thus
\begin{equation}
f^{ab\bc\bd} =\Gamma_A^{b\bc}(t^A)^{a\bd},
\end{equation}
where we have used the metric to raise an index.  Since
$f^{ab\bc\bd}=f^{*\bc\bd ab}$, we also see that the
$\Gamma_A^{a\bd}$ must be such that
\begin{equation}
f^{ab\bc\bd} =\sum_{AB} \Omega_{AB}(t^A)^{a\bd}(t^B)^{b\bc}
\end{equation}
for some real and symmetric $\Omega_{AB}$. If we now substitute this
expression into the fundamental identity, we find
\begin{equation}
0=\sum_{ABCDE}
\Omega_{CD}(c^{CB}{}_E\Omega_{AB}+c^{CB}{}_A\Omega_{EB})(t^A)^{a\bb}(t^E)^{c\bd}(t^D)^f{}_g,
\end{equation}
where the $c^{AB}{}_C$ are the structure constants of $\cal G$, \ie\
$[t^A,t^B]=c^{AB}{}_C t^C$. Defining $(j^A)^B{}_C=c^{AB}{}_C$ to be
the usual adjoint representation of $\cal G$, we see that the
fundamental identity implies
\begin{equation}
[\Omega,j^C]=0
\end{equation}
for all $C$, provided that $\Omega_{AB}$ is invertible. Thus by
Schur's Lemma, $\Omega_{AB}$ must be proportional to the identity in
each simple component of $\cal G$. In particular if the Lie algebra
${\cal G}$ is of the form
\begin{equation}
{\cal G}  = \oplus_\lambda {\cal G}_\lambda,
\end{equation}
where ${\cal G}_\lambda$ are commuting subalgebras of $\cal G$, then
we find
\begin{equation}\label{fequals}
f^{ab\bc\bd} =\sum_\lambda \omega_\lambda \sum_\alpha
(t^\alpha_\lambda)^{a\bd}(t^\alpha_\lambda)^{b\bc},
\end{equation}
where the $t^\alpha_\lambda$ span a $u(N)$ representation of the
generators of ${\cal G}_\lambda$ and the $\omega_\lambda$ are
arbitrary constants.

This would seem to furnish us with a very large class of ${\cal
N}=6$ Lagrangians. However, the $f^{ab\bc\bd}$ that we constructed in
(\ref{fequals}) do not necessarily satisfy $f^{ab\bc\bd}=-f^{ba\bc\bd}$.
This condition must be imposed by hand as an additional constraint.

This form for $f^{ab\bc\bd}$ allows us to write the `twisted'
Chern-Simons term as follows,
\begin{equation}\label{CS}
{\cal L}_{CS}= \sum_{\lambda} \frac{1}{4d_\lambda\omega_\lambda
}{\rm Tr}\left(\tilde A_{(\lambda)}\wedge d \tilde A_{(\lambda)}
+\frac{2}{3}\tilde A_{(\lambda)}\wedge \tilde A_{(\lambda)}\wedge
\tilde A_{(\lambda)}\right) .
\end{equation}
Here $\tilde A_{(\lambda)}{}^c{}_d=\tilde
A_{\mu\alpha}(t^\alpha_\lambda)^c{}_d dx^\mu$ is the projection of
the gauge field onto the eigenspace ${\cal G}_\lambda$ and
$d_\lambda$ is defined by the normalization ${\rm
Tr}(t_\lambda^\alpha
t_\lambda^\beta)=d_\lambda\delta^{\alpha\beta}$. We are free to
rescale the generators $t_\lambda^\alpha$ so that $d_k$ agrees with
the same quantity as calculated when the trace is taken to be in the
fundamental representation of ${\cal G}_\lambda$. For the path
integral to be well-defined, the coefficient of a Chern-Simons term
must be $k/4\pi$, where $k\in {\mathbb Z}$ \cite{Deser:1982vy}. This
leads to a quantization condition of the form $\omega_\lambda
=\pi/d_\lambda k$.

With these results, the Lagrangian can be written as
\begin{eqnarray}\label{niceaction}
\nonumber {\cal L} &=& -{\rm Tr}(D^\mu \bZ_A,D_\mu Z^A) -
i{\rm Tr}(\bar\psi^A,\gamma^\mu D_\mu\psi_A) -V+{\cal L}_{CS}\\[2mm]
&& -i{\rm Tr}(\bar\psi^A, [\psi_{A},
Z^B;\bZ_{B}])+2i{\rm Tr}(\bar\psi^A,[\psi_{B},Z^B;\bZ_{A}])\\
\nonumber &&+\frac{i}{2}\varepsilon_{ABCD}{\rm Tr}(\bar\psi^A,[
Z^C,Z^D;\psi^B]) -\frac{i}{2}\varepsilon^{ABCD}{\rm Tr}(\bZ_D,[\bar
\psi_{A},\psi_B;\bZ_{C}])\ ,
\end{eqnarray}
where
\begin{eqnarray}
  V &=& \frac{2}{3}{\rm Tr}(\Upsilon^{CD}_B,\bar\Upsilon^B_{CD}); \\
\nonumber  \Upsilon^{CD}_B &=&
  [Z^C,Z^D;\bZ_B]-\frac{1}{2}\delta^C_B[Z^E,Z^D;\bZ_E]+\frac{1}{2}\delta^D_B[Z^E,Z^C;\bZ_E]  ,
\end{eqnarray}
and ${\cal L}_{CS}$  is given in (\ref{CS}).

We close this section by constructing an infinite class of examples.  Let
$V_1$ and $V_2$ be complex vector spaces with
dimensions $N_1$ and $N_2$, respectively. Consider the vector space
$\cal A$ of linear maps $X$: $V_1\to V_2$. In general there is no
natural notion of a product on $\cal A$, but there is a natural
notion of a triple product:
\begin{equation}\label{ohhh}
[X,Y;\overline{Z}]=\lambda(XZ^\dagger Y -YZ^\dagger X).
\end{equation}
Here $\dagger$ denotes the transpose conjugate and $\lambda$ is an arbitrary
constant. If we introduce the inner product
\begin{equation}\label{ohhtrace}
{\rm Tr}(\overline{X},Y) = tr(X^\dagger Y),
\end{equation}
where $tr$ denotes the ordinary matrix trace, then one sees that
$f^{ab\bc\bd}$ satisfies the correct symmetry properties as well as
the fundamental identity.

{}From the Lie-algebra point of view, $V_1\cong {\mathbb C}^{N_1}$
and $V_2\cong {\mathbb C}^{N_2}$ can be regarded as the vector
space of the fundamental representation of $U(N_1)$ and $U(N_2)$
respectively. The maps $X$: $V_1\to V_2$ can then be viewed as
states in the bi-fundamental representation $({\bf N}_1,\bar {\bf
N}_2)$.  It is easy to see that the Lie algebra $\cal G$ acts on $X$
by
\begin{equation}
\delta X = XM_1-M_2^\dag X,
\end{equation}
where $M_1,M_2$ are elements of $u(N_1)$ and $u(N_2)$ respectively.
Thus we see that ${\cal G}=u(N_1)\oplus u(N_2)$. Finally one can
check that
\begin{equation}
\delta [X,Y;\bar Z] = [X,Y;\bar Z]M_1-M_2^\dag [X,Y;\bar Z],
\end{equation}
which is a manifestation of the fundamental identity.

With this choice of 3-algebra, the action (\ref{niceaction}) becomes
\begin{eqnarray}\label{ABJM}
\nonumber {\cal L} &=& -{\rm tr}(D^\mu Z_A^{\dag} D_\mu Z^A) -
i{\rm tr}(\bar\psi^{A\dag}\gamma^\mu D_\mu\psi_A) -V+{\cal L}_{CS}\\
\nonumber && -i\lambda{\rm tr}(\bar\psi^{A\dag} \psi_{A} Z^\dag_B
Z^B-\bar\psi^{A\dag} Z^B Z^\dag_B\psi_{A}
)\\
 &&+2i\lambda {\rm tr}(\bar\psi^{A\dag}\psi_{B} Z_A^\dag Z^B-\bar\psi^{A\dag} Z^B Z_A^\dag\psi_{B})\\
\nonumber &&+i\lambda\varepsilon_{ABCD}{\rm tr}(\bar\psi^{A\dag}
Z^C\psi^{B\dag} Z^D  ) -i\lambda\varepsilon^{ABCD}{\rm
tr}(Z_D^\dag\bar \psi_A Z_C^\dag\psi_B)\ .
\end{eqnarray}
For $N_1=N_2$ this is the ${\cal N}=6$ action of
\cite{Aharony:2008ug}, as written in component form in
\cite{Benna:2008zy}. For $N_1\ne N_2$ we obtain the $U(N_1)\times
U(N_2)$ models proposed in \cite{Aharony:2008gk}.

\section{\sl Conclusions}

In this paper we have studied the general form of
three-dimensional Lagrangians with ${\cal N}=6$ supersymmetry,
$SU(4)$ R-symmetry and a $U(1)$ global symmetry. The
resulting Lagrangians are of Chern-Simons form, with interacting
scalars and vectors that take values in a so-called 3-algebra.  As
with the ${\cal N}=8$ model previously constructed, the Lagrangian
is entirely determined by specifying a triple product on a 3-algebra
that satisfies the fundamental identity. For ${\cal N}=6$, the tensor
$f^{ab\bc\bd}$ that defines triple product need not be real or
totally antisymmetric.\footnote{In the special case that
$f^{abcd}$ is totally antisymmetric, it is also real and the
Lagrangian becomes that of the ${\cal N}=8$ theory.}

We believe that the ${\cal N}=6$ theories relevant for multiple
M2-branes are classified by tensors $f^{ab\bc\bd}$ that satisfy the
fundamental identity (\ref{FI}) and possess the symmetry properties
(\ref{Riemann}).  There is at least one very natural form for the
triple product that leads to the models of \cite{Aharony:2008ug}
with gauge group $U(N)\times U(N)$. It would certainly be
interesting to see if there are any other examples and hence other
models. For example ${\cal N}=6$ models with gauge group
$Sp(2N)\times O(2)$ have appeared in \cite{Hosomichi:2008jb}. In
addition, perhaps there is a connection to the embedding tensor
approach studied in \cite{Bergshoeff:2008cz}, or to the work of
\cite{FigueroaO'Farrill:2008zm,deMedeiros:2008bf} that classifies
totally antisymmetric 3-algebras.

We note that we have emphasized the role of triple products and
3-algebras even though the resulting Lagrangians can be viewed as
relatively familiar Chern-Simons gauge theories based on Lie
algebras with matter fields.  From our point of view, the dynamical
fields have interactions that are most naturally defined in terms of
a triple product.  Thus even though the 3-algebra may not be an
independent structure apart from a Lie algebra, we believe the
triple product is the central concept behind the M2-brane dynamics.
For example in \cite{Lambert:2008et}, the light states on the
Coulomb branch of the ${\cal N}=8$ theory were found to have masses,
at least in the classical theory, that are proportional to the area
of a triangle whose vertices end on the M2-branes. This is a
consequence of the appearance of the triple product in the dynamics
and hints to underlying M-theory degrees of freedom analogous to the
open strings that arise in D-branes.

\section*{\sl Acknowledgements}

We would like to thank the Institute for Advanced Study in
Princeton, where this work was initiated, and are grateful to O.\
Aharony, O.\ Bergman and J.\ Maldacena for discussions, as
well as informing us in advance of their work. We also thank P.\
Franche for comments, and the referee for questions
that led to improvement of sections 3 and 4.
J.B.\ is supported in part by the US National Science
Foundation grant NSF-PHY-0401513. N.L.\ is supported in part by the
PPARC grant PP/C507145/1 and the EU grant MRTN-CT-2004-512194.

\section*{\sl Appendix}

In this paper all spinor quantities in lower case letters are those
of three-dimensional Minkowski space with real two-component
spinors.  Spinor quantities with capitol letters refer to $11$-dimensional
Minkoswki space with 32 component spinors (although the
supersymmetry generators are always denoted by a lower case
$\epsilon$). In both cases $\gamma_\mu$, $\mu=0,1,2$ and $\Gamma_m$,
$m=0,1,2,...,10$ are sets of real $\gamma$-matrices with
$\gamma_{012}=1$ (resp. $\Gamma_{012345678910}=1$) and $\bar\epsilon
= \epsilon^T\gamma_0$ (resp. $\bar\epsilon = \epsilon^T\Gamma_0$ ).
The 8 transverse directions are labeled by the scalars $X^I$,
$I,J=1,..,8$ or in terms of 4 complex scalars $Z^A$, $A,B =1,2,3,4$,
with complex conjugates $\bZ_A$.

In three dimensions the Fierz transformation is
\begin{equation}
(\bar\lambda\chi)\psi = -\frac{1}{2}(\bar\lambda\psi)\chi
-\frac{1}{2} (\bar\lambda\gamma_\nu\psi)\gamma^\nu\chi .
\end{equation}
Furthermore, we note the following useful identities:
\begin{eqnarray}
\nonumber
\frac{1}{2}\bar\epsilon^{CD}_1\gamma_\nu\epsilon_{2CD}\,\delta^A_B&=&\bar\epsilon^{AC}_1\gamma_\nu\epsilon_{2BC}
-\bar\epsilon^{AC}_2\gamma_\nu\epsilon_{1BC}\\
 \nonumber
2\bar\epsilon^{AC}_1\epsilon_{2BD}
-2\bar\epsilon^{AC}_2\epsilon_{1BD}
&=&\bar\epsilon^{CE}_1\epsilon_{2DE}\delta^A_B
-\bar\epsilon^{CE}_2\epsilon_{1DE}\delta^A_B\\
\nonumber&-&\bar\epsilon^{AE}_1\epsilon_{2DE}\delta^C_B+\bar\epsilon^{AE}_2\epsilon_{1DE}\delta^C_B\\
&+&
\bar\epsilon^{AE}_1\epsilon_{2BE}\delta^C_D-\bar\epsilon^{AE}_2\epsilon_{1BE}\delta^C_D\\
\nonumber &-&
\bar\epsilon^{CE}_1\epsilon_{2BE}\delta^A_D+\bar\epsilon^{CE}_2\epsilon_{1DE}\delta^A_D\\
\nonumber \frac{1}{2}\varepsilon_{ABCD}
\bar\epsilon^{EF}_1\gamma_\mu\epsilon_{2EF}&=&\bar\epsilon_{1AB}\gamma_\mu\epsilon_{2CD}-\bar\epsilon_{2AB}\gamma_\mu\epsilon_{1CD}\\
&+&\bar\epsilon_{1AD}\gamma_\mu\epsilon_{2BC}-\bar\epsilon_{2AD}\gamma_\mu\epsilon_{1BC}\\
\nonumber&-&\bar\epsilon_{1BD}\gamma_\mu\epsilon_{2AC}+\bar\epsilon_{2BD}\gamma_\mu\epsilon_{1AC}.
\end{eqnarray}

In eleven dimensions the Fierz transformation is
\begin{eqnarray}\label{fierzplus}
 &&(\bar\epsilon_2\chi)\epsilon_1 -
(\bar\epsilon_1\chi)\epsilon_2 =\\
\nonumber&&-\frac{1}{16}\left(2(\bar\epsilon_2\Gamma_\mu\epsilon_1)\Gamma^\mu\chi
-(\bar\epsilon_2\Gamma_{IJ}\epsilon_1)\Gamma^{IJ}\chi
+\frac{1}{4!}(\bar\epsilon_2\Gamma_{\mu}\Gamma_{IJKL}\epsilon_1)\Gamma^\mu\Gamma^{IJKL}\chi
\right),
\end{eqnarray}
where  $\epsilon_1,\epsilon_2$ and $\chi$ have the same chirality
with respect to  $\Gamma_{012}$.

We also found the following identities useful:
\begin{eqnarray}
  \nonumber \Gamma_M\Gamma^{IJ}\Gamma^M&=& 4\Gamma^{IJ}\\
\nonumber  \Gamma_M\Gamma^{IJKL}\Gamma^M &=&0\\
\nonumber \Gamma^{IJP}\Gamma^{KLMN}\Gamma_P&=&-\Gamma^I\Gamma^{KLMN}\Gamma^J +\Gamma^J\Gamma^{KLMN}\Gamma^I \\
 \nonumber  \Gamma^I\Gamma^{KL}\Gamma^J - \Gamma^J\Gamma^{KL}\Gamma^I &=& 2\Gamma^{KL}\Gamma^{IJ}  - 2\Gamma^{KJ}\delta^{IL}
+ 2\Gamma^{KI}\delta^{JL}- 2\Gamma^{LI}\delta^{JK}\\
\nonumber&& + 2\Gamma^{LJ}\delta^{IK}- 4\delta^{KJ}\delta^{IL}+4\delta^{KI}\delta^{JL}   \\
  \nonumber \Gamma^{IJM}\Gamma^{KL}\Gamma_M &=&2\Gamma^{KL}\Gamma^{IJ}  - 6\Gamma^{KJ}\delta^{IL}
+ 6\Gamma^{KI}\delta^{JL}- 6\Gamma^{LI}\delta^{JK}\\
&& + 6\Gamma^{LJ}\delta^{IK}+
4\delta^{KJ}\delta^{IL}-4\delta^{KI}\delta^{JL}.
\end{eqnarray}

\end{document}